\documentclass[lettersize,journal]{IEEEtran}
\usepackage{amsmath,amssymb,amsfonts}
\usepackage{algorithm}
\usepackage{algpseudocode}
\usepackage{algorithmicx} 
\usepackage{calc}
\usepackage{array}
\usepackage[caption=false,font=normalsize,labelfont=sf,textfont=sf]{subfig}
\usepackage{textcomp}
\usepackage{stfloats}
\usepackage{url}
\usepackage{verbatim}
\usepackage{graphicx}
\usepackage{cite}
\usepackage{tabularx}
\usepackage{booktabs}
\usepackage{array}    
\usepackage{xcolor}
\usepackage{afterpage}
\usepackage{multirow}
\usepackage{bm}
\usepackage{tabularx} 
\usepackage{soul}
\usepackage{enumitem}
\usepackage{hyperref}
\usepackage{fontawesome5}
\usepackage{hhline}
\usepackage{siunitx}
\usepackage{caption}
\usepackage[most]{tcolorbox}
\usepackage{alltt}

\usepackage{pifont}

\definecolor{obsframe}{HTML}{6F8CBF} 
\definecolor{obsback}{HTML}{F5F6F8}  
\newtcolorbox{obsbox}{
enhanced,
colback=obsback,
colframe=obsframe,
boxrule=0.7pt,
arc=2mm,
left=1.5mm,
right=1.5mm,
top=0.8mm,
bottom=0.8mm,
fontupper=\bfseries\itshape,
height=8mm,
valign=center,
}

\definecolor{designframe}{HTML}{D0857F} 
\definecolor{designback}{HTML}{FDF4F5}  
\newtcolorbox{designbox}{
enhanced,
colback=designback,
colframe=designframe,
boxrule=0.7pt,
arc=2mm,
left=1.5mm,
right=1.5mm,
top=0.8mm,
bottom=0.8mm,
fontupper=\bfseries\itshape,
height=8mm,
valign=center,
}

\definecolor{noteframe}{HTML}{6F8CBF} 
\definecolor{noteback}{HTML}{F7F7F7}  
\newtcolorbox{notebox}{
enhanced,
colback=noteback,
colframe=black!55,
boxrule=0.5pt,
arc=2mm,
left=1.2mm,
right=1.2mm,
top=0.8mm,
bottom=0.8mm,
fontupper=\normalfont,
}

\definecolor{promptframe}{HTML}{B8B8B8}
\definecolor{promptblue}{HTML}{1F4E79}
\definecolor{promptred}{HTML}{FF1493}

\definecolor{stepone}{HTML}{F8CBAD}   
\definecolor{steptwo}{HTML}{C6E0B4}   
\definecolor{stepthree}{HTML}{BDD7EE} 

\newtcolorbox{promptbox}{
  enhanced,
  breakable,
  colback=white,
  colframe=promptframe,
  boxrule=0.35pt,
  arc=0pt,
  outer arc=0pt,
  sharp corners,
  width=\linewidth,
  left=1.5mm,
  right=1.5mm,
  top=1.2mm,
  bottom=1.2mm,
  fontupper=\ttfamily\scriptsize,
  before upper={
    \sloppy
    \setlength{\parindent}{0pt}
    \setlength{\parskip}{1.5pt}
    \raggedright
  },
}

\newcommand{\promptrole}[1]{%
{\color{promptblue}\bfseries\small - #1}\par
}

\newcommand{\phead}[1]{%
  {\color{promptred}\bfseries \# #1}\par
}

\newcommand{\hlone}[1]{{\sethlcolor{stepone}\hl{#1}}}
\newcommand{\hltwo}[1]{{\sethlcolor{steptwo}\hl{#1}}}
\newcommand{\hlthree}[1]{{\sethlcolor{stepthree}\hl{#1}}}

\newlength{\RequireLabelWidth}

\def\BibTeX{{\rm B\kern-.05em{\sc i\kern-.025em b}\kern-.08em
    T\kern-.1667em\lower.7ex\hbox{E}\kern-.125emX}}

\begin{document}

\title{Beyond Information Redundancy: Expanding Cross-Modal Knowledge Representation for Power Load Time Series Forecasting}

\author{Yuxuan Chen, Shuo Dai, Ruoyi Xu, Haipeng Xie \IEEEmembership{Senior Member, IEEE}
\thanks{All authors are with the National Key Laboratory for High Energy Pulsed Power, Xi’an Jiaotong University, Xi’an, Shaanxi, China. (corresponding author: H. Xie; e-mail: haipengxie@xjtu.edu.cn).}}

\markboth{IEEE Transactions on Emerging Topics in Computational Intelligence,~Vol.~xx, No.~x, August~2021}%
{xxx \MakeLowercase{\textit{et al.}}: Beyond Information Redundancy: Expanding Cross-Modal Knowledge Representation for Power Load Time Series Forecasting}


\maketitle

\begin{abstract}
Load forecasting is pivotal for stable power systems. Conventional uni-modal methods suffer from representation drift under data scarcity. While recent multi-modal approaches attempt to alleviate this, they exhibit severe information redundancy, merely recycling time series data via superficial intra-modal transformations. In this paper, we argue that the essence of multi-modal time series learning should expand representation manifolds via complementary cross-modal knowledge enrichment rather than duplicating redundant information, especially for few-shot scenarios prevalent in power systems. To this end, we propose KEMM-Net, a Knowledge-Enriched Multi-Modal Network for power load forecasting. KEMM-Net first constructs textual and visual embeddings to strengthen load time series representations from different knowledge perspectives. It then introduces a Partial Information Decomposition (PID)-guided cross-modal contrastive learning mechanism to achieve cross-modal semantic alignment and balance redundant, synergistic, and unique information for forecasting. Extensive experiments on real-world public datasets demonstrate that KEMM-Net consistently outperforms strong deep learning and multi-modal baselines, particularly in few-shot settings. Our code is available at https://anonymous.4open.science/r/KEMM-Net-2898
\end{abstract}


\vspace{-0.3em}
\section{Introduction}
\IEEEPARstart{E}{lectric} power is a fundamental energy source underpinning modern industry, economic growth, and daily life. Accurate power load forecasting is essential for optimizing resource allocation, improving unit commitment strategies, and ensuring the secure and stable operation of the power system~\cite{zhong2025value}.

Power load forecasting methods can be broadly categorized into statistical modeling and data-driven approaches. Statistical modeling primarily relies on autoregressive paradigms, such as ARIMA, exponential smoothing, which are often adept to solve short-term load forecasting~\cite{lopez2018parsimonious}. Although theoretically simple and lightweight, they depend heavily on data stationarity and clear trend or seasonal patterns, making them inadequate for capturing long-term dependencies and nonlinear dynamics in power load time series.

Data-driven approaches have become mainstream in recent years. Early methods centered on RNNs and CNNs, which are adept at capturing temporal dependencies and inter-variable relationships. With the advent of the Transformer architecture~\cite{vaswani2017attention}, attention-based models have advanced rapidly for time series forecasting, giving rise to numerous variants~\cite{wu2021autoformer,zhou2021informer,zhou2022fedformer,lin2024petformer} that have been widely applied to electric load forecasting~\cite{wang2023probabilistic}. However, despite the impressive results of Transformer approaches, their cross-entity generalization and performance in few-shot settings remain limited, making them ill-suited to the data-scarce scenarios common in industrial practice.

\begin{figure}[!t]
\centering
\includegraphics[width=\columnwidth]{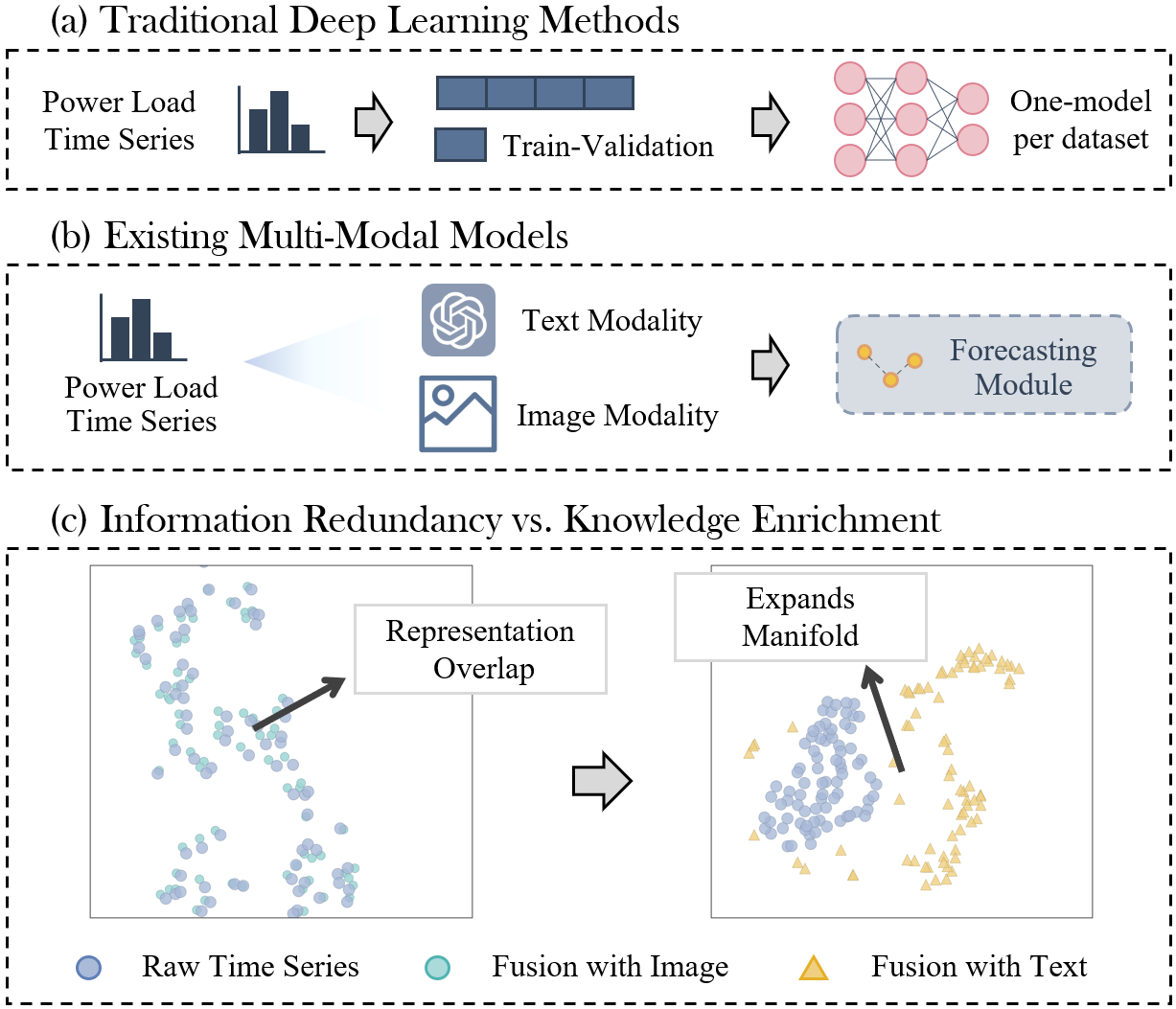}
\caption{(a) Traditional deep models suffer from one-model-per-dataset limitation; (b) Existing multi-modal approaches always reuse the knowledge extracted from time series; (c) Self-derived image modality leads to representation overlap. Text with domain knowledge expands the original manifold.}
\label{fig:1}
\vspace{-1.0em}
\end{figure}

More recently, researchers have explored auxiliary modalities for accurate forecasting. Specifically, LLM-based forecasting has been widely researched. Some approaches recast time series forecasting as prompt-based inference~\cite{xue2023promptcast,chen2025trace}, which bypasses large-scale retraining; however, pre-trained LLMs often struggle to capture fine-grained temporal dynamics. Other studies fine-tune or retrain parts of the encoder / decoder layers to encode numerical information~\cite{liu2024unitime,chang2025llm4ts,jintime}. Nevertheless, the inherent modality gap between numerical signals and text continues to limit performance.

Beyond text, researchers have further explored the multi-modal forecasting paradigm. Image modalities can explicitly preserve temporal dependencies and spatial correlations, and pretrained visual encoders have been adopted to extract informative embeddings for forecasting~\cite{chenvisionts,wang2025timemixer++}. Other studies unify text, image, or other modalities into a shared embedding space and design cross-modal fusion modules for prediction~\cite{zhongtime,liu2025timecma,chowdhury2026t3time,wuaurora}. These multi-modal approaches have improved forecasting accuracy and shown promising few-shot capability.

Despite this progress, a clear limitation remains in existing multi-modal frameworks: they often suffer from severe information redundancy, as the auxiliary modalities are inherently derived from the raw time series. This phenomenon is clearly evident in the multi-modal fused representations shown in Fig. 1(c). Fusing with self-derived image modalities results in a large representation overlap, whereas integrating cross-modal data with external domain prior knowledge significantly expands the original manifold to cover richer knowledge, offering a promising direction for few-shot forecasting. To achieve this, two key challenges need to be addressed: 1) how to obtain domain prior knowledge; 2) more importantly, how to enable the model to comprehend and align such multi-modal knowledge, without which the added modal information would degenerate into noise. In this context, we propose \textbf{KEMM-Net}, a \textbf{K}nowledge-\textbf{E}nriched \textbf{M}ulti-\textbf{M}odal \textbf{Net}work for power load forecasting. KEMM-Net consists of three key modules. The \textit{Multi-Modal Augment} module enriches raw time series representations through extra modalities. The \textit{Tri-Modality Encoding} module embeds all modalities into a joint space and performs forecasting with a cross-attention Transformer. The \textit{Multi-Modal Alignment} module further achieves domain-specific cross-modal semantic alignment via PID-guided contrastive learning.

Our contributions are summarized as follows:

\begin{itemize}[leftmargin=9pt]
  \item \textbf{Insight.} We identify a key limitation of existing multi-modal time series models in industrial few-shot forecasting: their bottleneck lies not merely in insufficient cross-modal fusion, but in the inherent limitation of recycling redundant information. To break this bottleneck, we demonstrate that enriching domain knowledge during the modal augmentation stage is a highly effective strategy.
  \item \textbf{Methodology.} We propose KEMM-Net, a Knowledge-Enriched Multi-Modal Network for power load forecasting. KEMM-Net introduces domain prior knowledge during multi-modal augmentation and further performs information-theoretic knowledge alignment to balance redundant, synergistic, and unique information for forecasting.
  \item \textbf{Evaluation.} We test on three distinct types of power load time series datasets. The results demonstrate that KEMM-Net consistently outperforms strong deep learning and multi-modal baselines, especially in few-shot scenarios.
\end{itemize}

\vspace{-0.3em}
\section{Related Works}
\subsection{Deep Learning-based Approaches}
This section focuses on deep learning-based approaches for time series forecasting. Early methodologies, including Recurrent Neural Networks (RNNs) and Convolutional Neural Networks (CNNs), demonstrated strong capabilities in capturing temporal and inter-channel dependencies. More recently, the field has been dominated by Transformer-based architectures. Seminal works in this area include Informer~\cite{zhou2021informer}, which introduced an efficient ProbSparse self-attention mechanism, Autoformer~\cite{wu2021autoformer}, which leveraged series decomposition and an auto-correlation mechanism, and FEDformer~\cite{zhou2022fedformer}, which innovatively applied attention in the frequency domain. More recent innovations like DLinear~\cite{zeng2023transformers} achieve efficient forecasting with a simple linear structure, while PatchTST~\cite{nietime} re-demonstrates the potential of Transformers through patch-based modeling and channel-independent design. These state-of-the-art methods have found widespread application in various industrial domains~\cite{jia2024two,wang2023probabilistic}. However, despite their significant success, these deep learning-based methods still exhibit limitations in cross-domain generalization and performance under data scarcity, making them difficult to adapt to the few-shot scenarios prevalent in industrial applications.

\vspace{-0.6em}
\subsection{Multi-Modal for Time Series}
The advent of Large Language Models (LLMs) has paved the way for text-enhanced forecasting. Approaches like PromptCast~\cite{xue2023promptcast} and TRACE~\cite{chen2025trace} formulated the task using natural language prompts. However, their efficacy was hindered by the intrinsic domain gap between numerical series and textual representations. To address this, GPT4TS~\cite{zhou2023one} and LLM4TS~\cite{chang2025llm4ts} re-trained tokenizers on time series data to enhance the LLM’s understanding of the numerical modality. Building on this, TimeLLM~\cite{jintime} and UniTime~\cite{liu2024unitime} leveraged pre-trained prior knowledge for successful few-shot forecasting. A key limitation, however, is that these methods primarily focused on refining time series embeddings rather than explicitly aligning representations across different modalities.

The integration of multi-modal analysis has recently emerged as a promising direction for time series forecasting. JLTSF~\cite{luo2026taming} aligns time series with the textual space and leverages language models to predict future trends. VisionTS~\cite{chenvisionts} employed a pre-trained visual autoencoder for zero-shot forecasting, however this method ignored the inter-variable interactions. TimeMixer++~\cite{wang2025timemixer++} utilized multi-scale frequency representations from images to improve the learning of temporal patterns. TimeCMA~\cite{liu2025timecma} first introduces a dual-branch paradigm that encodes text and time series separately and performs prediction through embedding fusion. Building on this line, T3Time~\cite{chowdhury2026t3time} further incorporates frequency-domain features, while TimeVLM~\cite{zhongtime} extends the framework to text, image, and time series modalities, and employs a frozen large vision-language model to obtain tri-modal embeddings. Nevertheless, these multi-modal methods still suffer from a fundamental limitation: their textual and visual modalities are largely extracted or transformed from the original time series, leading to repeated reuse of redundant information.

\vspace{0.5em}
\begin{notebox}
\textit{Remark:} 1) Recently, multi-modal time series foundation models have also emerged. Aurora~\cite{wuaurora} is trained on large-scale databases and can support cross-domain zero-shot inference. Such pre-trained large time series models follow a different paradigm from ours and are beyond the main scope of this paper; nevertheless, we include Aurora as a baseline in the few-shot experiments, with details provided in Section V.C. 2) Another line of work directly uses external information sources, such as news and policy reports, as additional modality inputs~\cite{hutowards,liu2024time}. In this study, we do not assume such external sources are always available, and instead focus on a more general setting where only historical time series are provided. Nevertheless, we further evaluate the compatibility of our approach with external-source information in Appendix D.
\end{notebox}

\section{Problem Formulation}
This study investigates the problem of power load forecasting, formulated as a classical time series prediction task. Let the historical load observations be represented as a sequence with $l$ time steps and $d$ channels:

\vspace{-0.5em}
\begin{equation}\label{eq:1}
{{X}_{1:l}}=\{{{x}_{1}},{{x}_{2}},...,{{x}_{l}}\}\in {{R}^{d\times l}}
\end{equation}

\noindent
where $x_t$ denotes the observation at time step $t$.

In addition, we perform multi-modal augmentation on the load time series to obtain auxiliary Text and Image modalities:

\vspace{-0.5em}
\begin{equation}\label{eq:2}
\begin{cases}
  T=LL{{M}_{\text{finetuned}}}({{X}_{1:l}}) \\ 
  I=BilinearInterp(Conv(FFT({{X}_{1:l}})))
\end{cases}
\end{equation}

Our goal is to learn an end-to-end function that takes $({{X}_{1:l}},T,I)$ as input and outputs the predicted load values for the next $h$ steps.

\vspace{-0.5em}
\begin{equation}\label{eq:3}
{{\hat{X}}_{l:l+h}}=\{{{\hat{x}}_{l}},{{\hat{x}}_{l+1}},...,{{\hat{x}}_{l+h}}|{{X}_{1:l}},T,I\}
\end{equation}

We consider forecasting tasks across multiple load scenarios, including user-level, building-level, and regional-level load forecasting within distribution networks. The prediction horizons cover both short-term and long-term scales.

\begin{figure*}[!t]
\centering
\includegraphics[width=\textwidth]{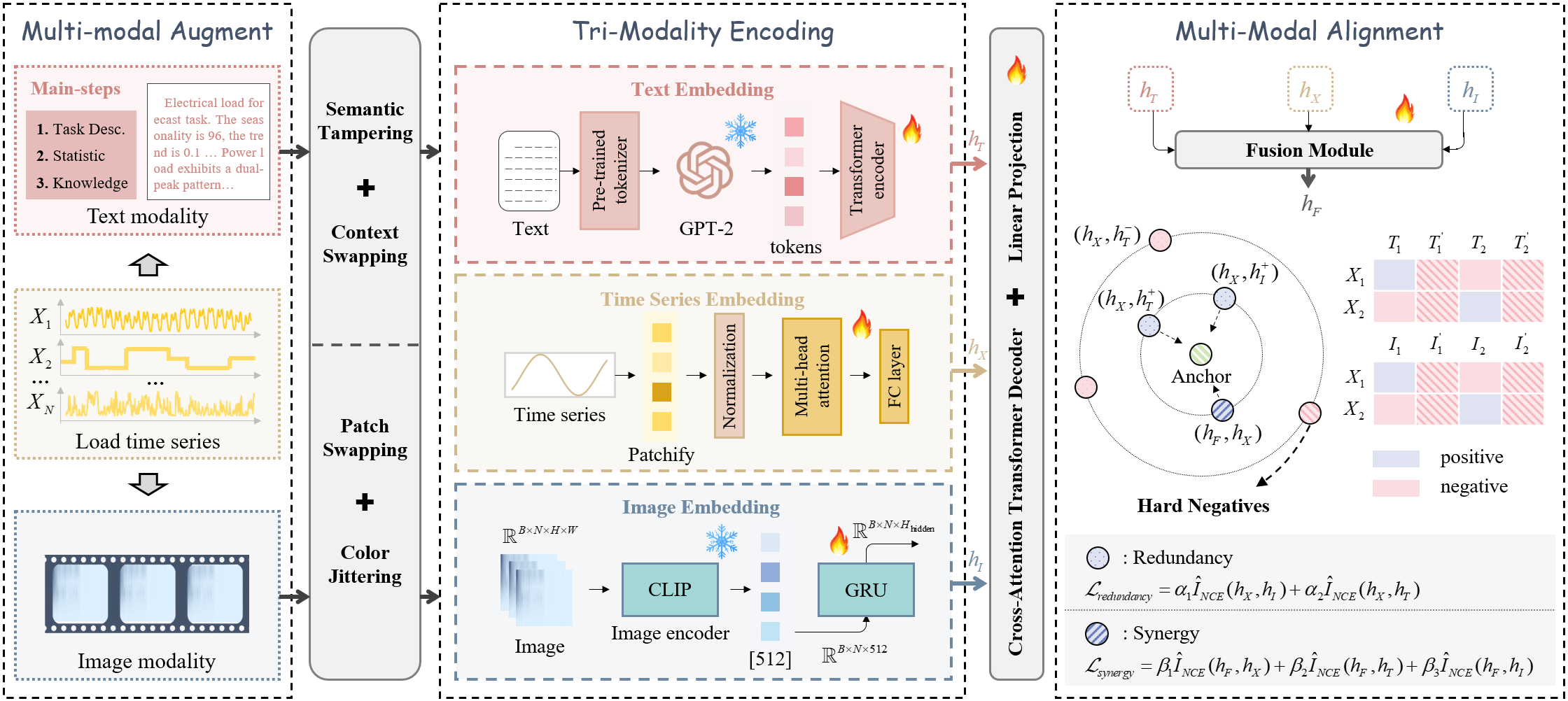}
\caption{The overall framework of KEMM-Net}
\label{fig:2}
\end{figure*}

\section{Methodology}
The overall framework of the proposed model, KEMM-Net, is shown in Fig. 2. KEMM-Net comprises three components:

\begin{itemize}[leftmargin=9pt]
  \item \textbf{Multi-modal Augment.} The raw load time series is transformed into an image modality to preserve temporal patterns, while an instruction-finetuned LLM generates the text modality with domain prior knowledge to enrich time series representations.
  \item \textbf{Tri-Modality Encoding.} We obtain embeddings for all three modalities and then employ a cross-attention Transformer as the forecaster.
  \item \textbf{Multi-modal Alignment.} A PID-guided multi-modal contrastive learning paradigm aligns domain-specific knowledge across modalities and balances redundant, synergistic, and unique information for forecasting.
\end{itemize}

\subsection{Multi-Modal Augment}

\noindent
\textbf{Text Modality.} To enrich load time series with domain-aware semantic cues, we use an instruction-finetuned LLM to generate structured textual descriptions for each load segment. The LLM follows a fixed output template (Fig.~3, left), ensuring consistent textual inputs across samples. The resulting textual view consists of three parts:

\vspace{-0.5em}
\begin{equation}\label{eq:4}
T={{T}_{\text{task}}}+{{T}_{\text{stat}}}+{{T}_{\text{knowledge}}}
\end{equation}

\noindent
Here, ${T}_{\text{task}}$ gives a brief description of the forecasting task (e.g., subject, resolution, and other metadata); ${T}_{\text{stat}}$ summarizes statistics of the segment (seasonality, trend, min/max, mean, kurtosis, skewness, etc.); and ${T}_{\text{knowledge}}$ encodes domain knowledge, such as load-specific patterns leveraged from the LLM’s pretrained priors knowledge to provide additional cues for downstream forecasting. The detail of this instruction paradigm can be found in Appendix D.

\vspace{0.3em}
\noindent
\textbf{Image Modality.} We first extract frequency features via the Fast Fourier Transform (FFT):

\begin{equation}\label{eq:5}
FFT({{X}_{1:l}})[k]=\sum\nolimits_{t=0}^{l-1}{{{x}_{t}}\cdot {{e}^{-2\pi ikt/l}}}
\end{equation}

\noindent
where $k$ is the (discrete) frequency index.

We then concatenate the FFT features with the original series and pass them through one 1-D convolution and two 2-D convolutions to obtain a reshaped feature map. To match the image size typically seen during CLIP pretraining, we resize it with bilinear interpolation:

\vspace{-0.5em}
\begin{equation}\label{eq:6}
I(u,v)= \sum_{i=0}^{1}\sum_{j=0}^{1}w_{ij}\, I(x_i,y_j)
\end{equation}

\noindent
where $(x_i,y_j)$ are four neighboring integer coordinates of $(u,v)$, and $w_{ij}$ are determined on relative distances.

After resizing, we normalize values to [0,255] and obtain the image modality $I\in {{\mathbb{R}}^{B\times H\times W}}$ (we set $H\times W=224\times 224$)

\vspace{0.5em}
\noindent
\textbf{Negatives construction.} Because different modalities exhibit substantial domain gaps, naive feature alignment may cause negative transfer. We therefore perform semantic alignment via contrastive learning, which requires abundant positive/negative pairs. For text, we construct negatives by:

\begin{itemize}[leftmargin=9pt]
  \item Context swapping: replace metadata in the description (e.g., turn a user-level sequence into “Distribution network …”).
  \item Semantic tampering: corrupt numeric statements (e.g., alter seasonality to “The seasonality is $<$22$>$”).
\end{itemize}

For images, we construct negatives by:

\begin{itemize}[leftmargin=9pt]
  \item Patch swapping: crop an image and swap its patches.
  \item Color jittering: randomly perturb brightness, contrast, and saturation.
\end{itemize}

\begin{figure*}[!t]
\centering
\includegraphics[width=.95\textwidth]{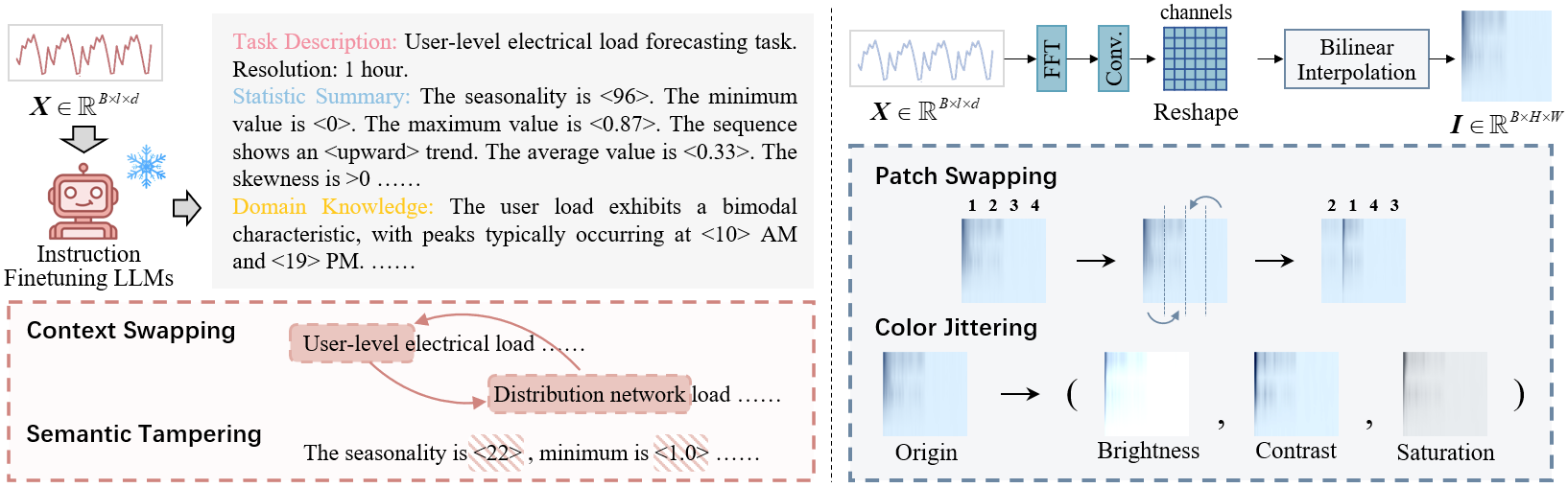}
\caption{Multi-modal augment and negatives construction}
\label{fig:3}
\end{figure*}

\subsection{Tri-Modality Encoding}

\noindent
\textbf{Time Series Encoding.} We first patchify the load sequence and apply normalization. The normalized patches are then fed into a multi-head self-attention layer $Att{{n}_{\text{MH-self}}}(\cdot )$ with a residual connection to capture temporal dependencies:

\vspace{-0.5em}
\begin{equation}\label{eq:7}
Att{{n}_{\text{MH-self}}}({{\bar{X}}^{i}})=Attention({{\omega }_{q}}{{\bar{X}}^{i}},{{\omega }_{k}}{{\bar{X}}^{i}},{{\omega }_{v}}{{\bar{X}}^{i}})
\end{equation}

\vspace{-0.5em}
\begin{equation}\label{eq:8}
{{\tilde{X}}^{i}}=Att{{n}_{\text{MH-self}}}({{\bar{X}}^{i}})+{{\bar{X}}^{i}}
\end{equation}

\noindent
Here, ${{\bar{X}}^{i}}$ denotes the $i$-th normalized patch segment, and ${{\omega }_{q}}$, ${{\omega }_{k}}$, ${{\omega }_{v}}$ are learnable linear projections.

Finally, a feed-forward projection produces the time series embedding:

\vspace{-0.5em}
\begin{equation}\label{eq:9}
{{h}_{X}}=FCL({{\tilde{X}}^{i}})
\end{equation}

\noindent
\textbf{Text Encoding.} We use a frozen GPT-2 to tokenize and represent the raw text. GPT-2 is a Transformer decoder with causal multi-head self-attention, which enforces that each token attends only to previous tokens and itself:

\vspace{-0.5em}
\begin{equation}\label{eq:10}
{{\bar{T}}^{i}}=Att{{n}_{\text{MMH-self}}}({{T}^{i}})+{{T}^{i}}
\end{equation}

\vspace{-0.5em}
\begin{equation}\label{eq:11}
{{\tilde{T}}^{i}}=FFN({{\bar{T}}^{i}})+{{\bar{T}}^{i}}
\end{equation}

We then attach a trainable Transformer-based encoder to adapt the textual features to the forecasting task:

\vspace{-0.5em}
\begin{equation}\label{eq:12}
{{h}_{T}}=Encode{{r}_{\text{prompt}}}({{\tilde{T}}^{i}})
\end{equation}

\noindent
\textbf{Image Encoding.} Prior studies~\cite{zhongtime,chenvisionts} treat the image corresponding to each segment as an independent input, failing to capture potential temporal dependencies across segments. We follow a standard temporal consistency design from video modeling: a group of $N$ images $I\in {{\mathbb{R}}^{B\times N\times H\times W}}$ is fed to a pretrained CLIP image encoder to obtain frame-wise embeddings ${{Z}_{I}}\in {{\mathbb{R}}^{B\times N\times 512}}$. We then attach a trainable GRU module to model within-group dependencies and produce the image representation ${{h}_{I}}\in {{\mathbb{R}}^{B\times N\times {{H}_{\text{hidden}}}}}$ :

\vspace{-0.5em}
\begin{equation}\label{eq:13}
h_{t}^{I}=GRU({{z}_{t}},h_{t-1}^{I}),\text{ }{{h}_{I}}=[h_{1}^{I},...,h_{N}^{I}]
\end{equation}

\noindent
\textbf{Forecaster.} We adopt a cross-modal multi-head attention decoder followed by a linear projection head. The time series embedding serves as queries, while the fused text-image representation provides the keys and values, yielding the final multi-step predictions.

\subsection{Multi-Modal Alignment}

Different modalities provide distinct knowledge perspectives for load forecasting. To make these heterogeneous knowledge sources jointly serve forecasting, a dedicated cross-modal semantic alignment mechanism is required.

Our design is inspired by partial information decomposition. PID posits that the information about a target can be decomposed into contributions induced by different factors~\cite{williams2010nonnegative}. We argue that the integrated gains achieved through interactions among modalities naturally accord with the PID perspective.

Let $\{{{M}_{1}},{{M}_{2}},...,{{M}_{n}}\}$ denote a set of modalities, $Y$ the prediction target, and $Info(\cdot )$ the informative content provided by the n-modal ensemble. Then $Info(Y;{{M}_{1}},{{M}_{2}},...,{{M}_{n}})$ can be viewed as the aggregate of three interaction types: \textbf{uniqueness}, \textbf{redundancy}, and \textbf{synergy}.

1) Uniqueness. Modality ${{M}_{i}}$ may contribute information that no other modality ${{M}_{j}},i\ne j$ provides. 
Let ${{U}_{i}}$ denote the unique contribution of modality $i$.

2) Redundancy. Two modalities ${{M}_{i}}$ and ${{M}_{j}}$ can carry the same or overlapping information. 
We use $R$ to denote the redundant contribution.

3) Synergy. The joint use of ${{M}_{i}}$ and ${{M}_{j}}$ can reveal additional information about $Y$ that is unavailable from any single modality. We denote this cross-modal synergistic contribution by $S$. In the three-variable case, the interaction diagram in Fig. 4 illustrates that synergy is not the set-theoretic union of modalities; rather, it corresponds to genuinely novel information that emerges only from their combination.

\begin{figure}[h]
\vspace{-0.3em}
\centering
\includegraphics[width=0.9\columnwidth]{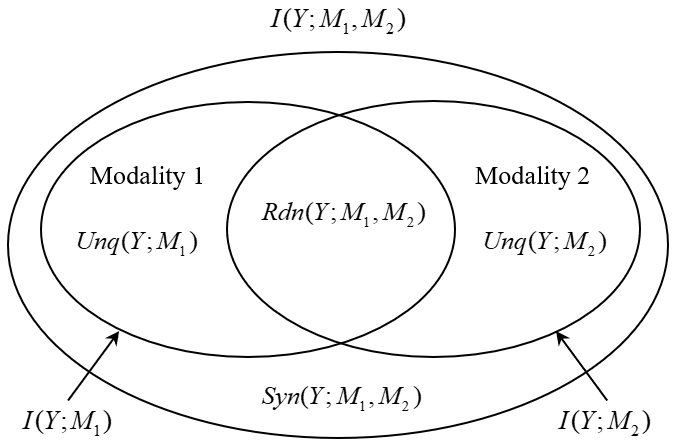}
\caption{PID for three-variable system}
\label{fig:x}
\vspace{-0.3em}
\end{figure}

\vspace{-0.3em}
In KEMM-Net, we consider three modalities—time series, text, and image—together with the prediction target $Y$, forming a \textbf{4}-variable PID system. Compared with 3-variable case~\cite{dufumieralign}, the 4-variable PID is more involved and introduces entangled information terms. The 4-variable PID CL has not been addressed~\cite{dufumieralign}, but in our scenario we can simplify the standard 4-variable system based on the derivation below.

\vspace{0.5em}
\noindent
\textbf{Lemma 1:} The information provided by a single modality and by a set of modalities can be expressed as combinations of redundancy $R$, uniqueness $U$, and synergy $S$:

\begin{equation}\label{eq:x}
\begin{cases}
  Info(Y;M_{i})=R+U_{i} \\ 
  Info(Y;M_{1},...,M_{n})=R+S+\sum{U_{i}}
\end{cases}
\end{equation}

\noindent
\textbf{Definition 1:} When PID involves more than two modalities, the system becomes complex and \textit{entangled information terms} arise. We define the entangled information between modalities $i$ and $j$ as $E{{I}_{ij}}$, which aggregates all cross-terms mixing redundancy and synergy related to $i$ and $j$ (e.g., the redundancy shared by the pairwise synergies \{X,T\} and \{X,I\}, denoted \{X,T\}\{X,I\}; see Appendix A, Fig. A2)

\vspace{0.5em}
\noindent
\textbf{Corollary 1:} In 4-variable system ${{M}_{X}}$, ${{M}_{T}}$, ${{M}_{I}}$, $Y$, the information provided by different modality combinations admits the following simplifications (Proof is given in Appendix A)

\vspace{-0.5em}
\begin{equation}\label{eq:x}
\begin{cases}
  Info(Y;M_{X},M_{T})=S_{XT}+U_{T}+Info(Y;M_{X})+EI_{XT} \\ 
  Info(Y;M_{X},M_{I})=S_{XI}+U_{I}+Info(Y;M_{X})+EI_{XI} \\ 
  \begin{aligned}
    Info(Y;M_{X},M_{T},M_{I})={} & S_{XTI}+S_{XI}+S_{XT}+EI_{XTI} \\
    & +Info(Y;M_{X})+U_{T}+U_{I}
  \end{aligned}
\end{cases}
\end{equation}

\noindent
\textbf{Assumption 1:} Let ${{f}_{\theta }}(\cdot )$ denote the embedding function of a modality. We assume that, once the model is ideally trained, the embedding $f_{\theta }^{*}$ preserves all the information carried by that modality. Formally,

\begin{equation}\label{eq:x}
Info(Y;f_{\theta }^{*}(M))\approx Info(Y;M)
\end{equation}

\noindent
\textbf{Lemma 2:} Given Assumption 1 and Lemma 1, we obtain:

\begin{equation}\label{eq:x}
Info(Y;f_{\theta }^{*}({{M}_{i}}))\approx R+{{U}_{i}}
\end{equation}

\noindent
\textbf{Assumption 2.} The auxiliary modalities $T$ and $I$, together with their negatives $T'$ and $I'$, have the capacity to collaborate with $X$ to accomplish task $Y$. This is a mild assumption: even in unimodal settings it holds; otherwise the total available information would be insufficient for prediction.

\vspace{0.5em}
\noindent
\textbf{Corollary 2.} Under Assumption 2, when training is adequate, the fused information can approximate the target $Y$. Combined with Lemma 2, we obtain:

\begin{equation}
\begin{aligned}
  & \phantom{{}\approx{}} Info(h_{F};f_{\theta}^{*}(M_{X}),f_{\theta}^{*}(M_{T}),f_{\theta}^{*}(M_{I})) \\
  & \approx Info(Y;f_{\theta}^{*}(M_{X}),f_{\theta}^{*}(M_{T}),f_{\theta}^{*}(M_{I})) \\
  & \approx Info(Y;X,T,I)
\end{aligned}
\end{equation}

\noindent
where $h_F$ is the fused representation. The first approximation follows from Assumption 2 and the second from Lemma 2.

From the expressions above and Corollary 1, the last term aggregates all synergistic information. Meanwhile, by Lemma 2, the information contributed by any single modality decomposes into redundancy and uniqueness. This motivates our design: we adopt the classic \textit{InfoNCE} loss to measure similarity between positive/negative pairs. Cross-modal losses capture redundancy $R$, while losses between the fused representation and each unimodal representation capture synergy $S$:

\vspace{-0.5em}
\begin{equation}\label{eq:x}
{\small
\begin{cases}
  {{\mathcal{L}}_{rdn}}={{\alpha }_{1}}{{{\hat{I}}}_{NCE}}({{h}_{X}},{{h}_{T}})+{{\alpha }_{2}}{{{\hat{I}}}_{NCE}}({{h}_{X}},{{h}_{I}}) \\ 
  {{\mathcal{L}}_{syn}}={{\beta }_{1}}{{{\hat{I}}}_{NCE}}({{h}_{F}},{{h}_{X}})+{{\beta }_{2}}{{{\hat{I}}}_{NCE}}({{h}_{F}},{{h}_{T}})+{{\beta }_{3}}{{{\hat{I}}}_{NCE}}({{h}_{F}},{{h}_{I}})
\end{cases}
}
\end{equation}

The meaning of Eq. (19) is straightforward: ${{\mathcal{L}}_{rdn}}$ pulls positive  pairs together, encouraging the model to learn shared, consistent signals. ${{\mathcal{L}}_{syn}}$ encourages cross-modal knowledge to contribute to the final fusion; if a modality (e.g., text) helps align the fused representation with others, the model tends to preserve and exploit it, reflecting multi-modal synergy.

Finally, we regard uniqueness $U$ as crucial for downstream forecasting. Rather than optimizing it directly with a contrastive loss, we preserve it indirectly via the task loss, as ${{\mathcal{L}}_{prediction}}$ forces the model to retain modality-specific cues that benefit prediction. Our overall objective is:

\vspace{-0.5em}
\begin{equation}\label{eq:x}
{{\mathcal{L}}_{total}}={{\mathcal{L}}_{prediction}}+{{\lambda }_{1}}{{\mathcal{L}}_{rdn}}+{{\lambda }_{2}}{{\mathcal{L}}_{syn}}
\end{equation}

\vspace{0.5em}
\begin{notebox}
\textit{Note:} In our knowledge enrichment context, synergy is defined slightly differently from~\cite{dufumieralign}. We focus on the complementary effect of knowledge itself, rather than the contribution of cross-modal fusion to a learning system. Moreover, redundancy should not be entirely suppressed, as it helps capture typical temporal patterns. The key is therefore not to eliminate redundancy, but to regulate different information roles toward forecasting-oriented representation learning.
\end{notebox}

\section{Experiments}

In this section, we evaluate KEMM-Net on various datasets and address three questions: 1) Does multi-modal information enable accurate power load forecasting? 2) Does KEMM-Net remain effective under few-shot settings? 3) Can KEMM-Net expand time series representations with richer knowledge coverage and translate such enrichment into accurate forecasting?

\subsection{Experiment Setup}
\noindent
\textbf{Datasets.} We use three real-world datasets, Electricity\footnote{\url{https://archive.ics.uci.edu/ml/datasets/ElectricityLoadDiagrams20112014}}, NEST~\cite{heer2024comprehensive},  StoreNet~\cite{trivedi2024comprehensive}, covering user, building, and distribution network loads. Details are provided in Appendix B.

\vspace{0.3em}
\noindent
\textbf{Baselines.} We compare KEMM-Net with strong baselines: 1) Multi-modal models: Time-LLM~\cite{jintime}, PromptCast~\cite{xue2023promptcast}, T3Time~\cite{chowdhury2026t3time}, TimeVLM~\cite{zhongtime} ,VisionTS~\cite{chenvisionts}, Aurora~\cite{wuaurora}. 2) Dlinear~\cite{zeng2023transformers}. 3) Transformer models: PatchTST~\cite{nietime}, FEDformer~\cite{zhou2022fedformer}, Autoformer~\cite{wu2021autoformer}. We evaluate predictive performance using two common metrics: MAE and MSE.

\vspace{0.3em}
\noindent
\textbf{Implementation Details.} The multi-modal embedding dimension is set to 128, the input window length is 480 and forecasting horizons are $h\in \{24,96,192,336\}$. We use a batch size of 32, an initial learning rate of 0.0001, the Adam optimizer, and train for 10 epochs with early stopping (patience = 5) and dropout = 0.1. For other deep baselines, we tune training hyperparameters while keeping their model architectures unchanged to ensure a fair comparison. The forecaster is trained with only MSE loss for the first 20\% epochs, after which contrastive losses are introduced. All experiments are conducted on Intel Xeon Gold 6254 processor, NVIDIA A800 GPU.

\begin{table*}[t]
\centering
\caption{Main results. best results are in \textbf{bold}, and the second best is \underline{underlined}.}
\label{tab:main results}
\renewcommand{\arraystretch}{1.2}  
\resizebox{\textwidth}{!}{
\begin{tabular}{c|c||cc|cc|cc|cc|cc|cc||cc|cc|cc|cc}
\toprule
\multicolumn{2}{c||}{\multirow{2}{*}{Methods}} & \multicolumn{12}{c||}{\includegraphics[height=1.0em]{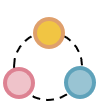}  \textit{\textbf{Multi-Modal Models}}} & \multicolumn{8}{c}{\includegraphics[height=1.0em]{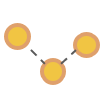}  \textit{\textbf{Deep Models}}} \\
\cline{3-22}
\multicolumn{2}{c||}{} & \multicolumn{2}{c|}{KEMM-Net} & \multicolumn{2}{c|}{Time-LLM} & \multicolumn{2}{c|}{PromptCast} & \multicolumn{2}{c|}{T3Time} & \multicolumn{2}{c|}{TimeVLM} & \multicolumn{2}{c||}{VisionTS} & \multicolumn{2}{c|}{Dlinear} & \multicolumn{2}{c|}{PatchTST} & \multicolumn{2}{c|}{FEDformer} & \multicolumn{2}{c}{Autoformer} \\
\cline{3-22}
\multicolumn{2}{c||}{} & MSE & MAE & MSE & MAE & MSE & MAE & MSE & MAE & MSE & MAE & MSE & MAE & MSE & MAE & MSE & MAE & MSE & MAE & MSE & MAE \\
\midrule
\multirow{5}{*}{\rotatebox{90}{StoreNet}} 
& 24 & \underline{0.521} & \underline{0.364} & 0.583 & 0.408 & 0.563 & 0.391 & 0.534 & 0.382 & 0.531 & 0.387 & 0.601 & 0.439 & 0.534 & 0.374 & \textbf{0.519} & \textbf{0.352} & 0.605 & 0.494 & 0.599 & 0.465 \\
& 96 & \textbf{0.550} & \textbf{0.374} & 0.585 & 0.434 & 0.578 & 0.424 & 0.599 & 0.411 & 0.575 & 0.409 & 0.668 & 0.479 & \underline{0.561} & 0.407 & 0.581 & \underline{0.375} & 0.671 & 0.537 & 0.658 & 0.516 \\
& 192 & \textbf{0.573} & \textbf{0.395} & 0.599 & 0.434 & \underline{0.590} & 0.443 & 0.630 & 0.418 & 0.599 & 0.440 & 0.675 & 0.462 & 0.596 & 0.453 & 0.603 & \underline{0.395} & 0.691 & 0.574 & 0.664 & 0.492 \\
& 336 & \textbf{0.595} & \textbf{0.403} & 0.622 & 0.448 & 0.611 & 0.471 & 0.658 & \underline{0.422} & 0.634 & 0.467 & 0.685 & 0.472 & \underline{0.604} & 0.446 & 0.699 & 0.428 & 0.703 & 0.545 & 0.717 & 0.541 \\
\cline{2-22}
& Avg & \textbf{0.560} & \textbf{0.384} & 0.597 & 0.431 & 0.586 & 0.432 & 0.605 & 0.408 & 0.585 & 0.426 & 0.657 & 0.463 & \underline{0.574} & 0.420 & 0.601 & \underline{0.385} & 0.668 & 0.537 & 0.660 & 0.503 \\
\midrule
\multirow{5}{*}{\rotatebox{90}{NEST}} 
& 24 & \underline{0.094} & \underline{0.174} & 0.155 & 0.250 & 0.139 & 0.212 & 0.115 & 0.209 & 0.129 & 0.233 & 0.223 & 0.316 & 0.096 & 0.176 & \textbf{0.090} & \textbf{0.170} & 0.227 & 0.343 & 0.247 & 0.362 \\
& 96 & \textbf{0.155} & \textbf{0.231} & 0.177 & 0.259 & 0.169 & 0.260 & 0.177 & 0.272 & 0.159 & 0.251 & 0.354 & 0.424 & \underline{0.157} & \underline{0.247} & 0.165 & 0.251 & 0.263 & 0.375 & 0.365 & 0.443 \\
& 192 & \textbf{0.197} & \textbf{0.269} & 0.206 & 0.278 & 0.211 & 0.297 & 0.215 & 0.285 & 0.204 & \underline{0.277} & 0.422 & 0.465 & \underline{0.198} & 0.288 & 0.212 & 0.287 & 0.321 & 0.411 & 0.343 & 0.430 \\
& 336 & \textbf{0.231} & \textbf{0.300} & 0.247 & 0.313 & 0.256 & 0.359 & 0.236 & 0.319 & 0.250 & 0.315 & 0.450 & 0.478 & \underline{0.234} & 0.319 & 0.239 & \underline{0.312} & 0.297 & 0.389 & 0.526 & 0.541 \\
\cline{2-22}
& Avg & \textbf{0.169} & \textbf{0.243} & 0.196 & 0.275 & 0.194 & 0.282 & 0.186 & 0.271 & 0.186 & 0.269 & 0.362 & 0.421 & \underline{0.171} & 0.257 & 0.177 & \underline{0.255} & 0.277 & 0.380 & 0.370 & 0.444 \\
\midrule
\multirow{5}{*}{\rotatebox{90}{Electricity}} 
& 24 & 0.113 & 0.221 & 0.158 & 0.273 & 0.119 & 0.210 & 0.130 & 0.244 & 0.128 & 0.239 & 0.200 & 0.310 & \underline{0.109} & \underline{0.207} & \textbf{0.103} & \textbf{0.207} & 0.185 & 0.304 & 0.182 & 0.302 \\
& 96 & \textbf{0.140} & \textbf{0.241} & 0.162 & 0.272 & 0.165 & 0.255 & 0.172 & 0.282 & 0.143 & 0.245 & 0.243 & 0.345 & 0.146 & \underline{0.248} & \underline{0.143} & 0.249 & 0.227 & 0.337 & 0.222 & 0.332 \\
& 192 & \textbf{0.155} & \textbf{0.255} & 0.182 & 0.286 & 0.187 & 0.285 & 0.251 & 0.361 & 0.160 & 0.266 & 0.269 & 0.367 & \underline{0.159} & \underline{0.260} & 0.165 & 0.262 & 0.251 & 0.355 & 0.230 & 0.339 \\
& 336 & \textbf{0.186} & \textbf{0.287} & 0.211 & 0.320 & 0.208 & 0.323 & 0.234 & 0.339 & 0.199 & 0.304 & 0.298 & 0.409 & \underline{0.191} & \underline{0.294} & 0.194 & 0.299 & 0.241 & 0.352 & 0.233 & 0.344 \\
\cline{2-22}
& Avg & \textbf{0.148} & \textbf{0.251} & 0.178 & 0.288 & 0.170 & 0.268 & 0.197 & 0.306 & 0.158 & 0.264 & 0.252 & 0.358 & 0.151 & \underline{0.252} & \underline{0.151} & 0.254 & 0.226 & 0.337 & 0.217 & 0.329 \\
\bottomrule
\end{tabular}
}
\vspace{-0.5em}
\end{table*}

\subsection{Main Results}
We tested across multiple scales $\{24,96,192,336\}$, covering both short-term and long-term forecasting needs, as shown in Table \ref{tab:main results}. It is evident that the proposed model KEMM-Net achieves the best average performance across the three datasets. Compared to the second best model, KEMM-Net reduces MAE and MSE errors by 4.7\% and 6.8\%, respectively.

\begin{itemize}[leftmargin=9pt]
  \item Among deep models using single modalities, PatchTST achieves relatively better predictions, validating the effectiveness patch-wise temporal modeling in our model design.
  \item Among multi-modal models, PromptCast directly verbalizes time series as raw text and yields limited accuracy, whereas Time-LLM improves performance by adapting time series embeddings to language models. This indicates that effective forecasting requires more than a simple modality conversion. Additionally, VisionTS, pretrained on full domain data, shows lower prediction accuracy than Time-VLM which uses modality-specific fine-tuning. Our proposed KEMM-Net enriches load representations during multi-modal augmentation and further aligns heterogeneous semantics through PID-guided contrastive learning, leading to the most accurate predictions.
\end{itemize}

We compared the training/inference time and GPU memory usage of KEMM-Net with other multi-modal baselines, as shown in Table 
\ref{tab:training time}. Despite integrating three modalities, KEMM-Net achieves the lowest training time (including LLM inference), since we use frozen main encoders (GPT-2, CLIP) and trained a small amount of parameters for domain knowledge alignment. VisionTS, which uses a fully pretrained VAE, has relatively low overall inference time but struggles to guarantee prediction accuracy. KEMM-Net strikes a good trade-off, making it suitable for deployment in real-world industrial scenarios.

\begin{table}[t]
\centering
\caption{Computational efficiency analysis}
\label{tab:training time}
\renewcommand{\arraystretch}{0.95}
\resizebox{\columnwidth}{!}{
\begin{tabular}{l|c|c|c}
\toprule
{} & \textbf{Training Time (min)} & \textbf{Inference Time (s)} & \textbf{GPU Mem.} \\
\midrule
KEMM-Net & 22.7 & 3.9 & 1508M \\
Time-LLM & 264.0 & 12.5 & 11228M \\
T3Time & 183.4 & 4.6 & 814M \\
TimeVLM & 41.5 & 4.5 & 1389M \\
VisionTS & 26.2 & 3.3 & 3074M \\
\bottomrule
\end{tabular}
}
\vspace{-1.0em}
\end{table}

\vspace{-0.3em}
\subsection{Few-Shot Scenarios}
In industrial applications, the increasing number of newly integrated entities often leads to limited historical data for each entity, creating numerous few-shot scenarios. Traditional deep models typically struggle in such few-shot settings, as they are prone to overfitting on scarce data. Therefore, it is crucial to test the model's few-shot learning capabilities.

\begin{figure*}[!t]
\centering
\includegraphics[width=0.95\textwidth]{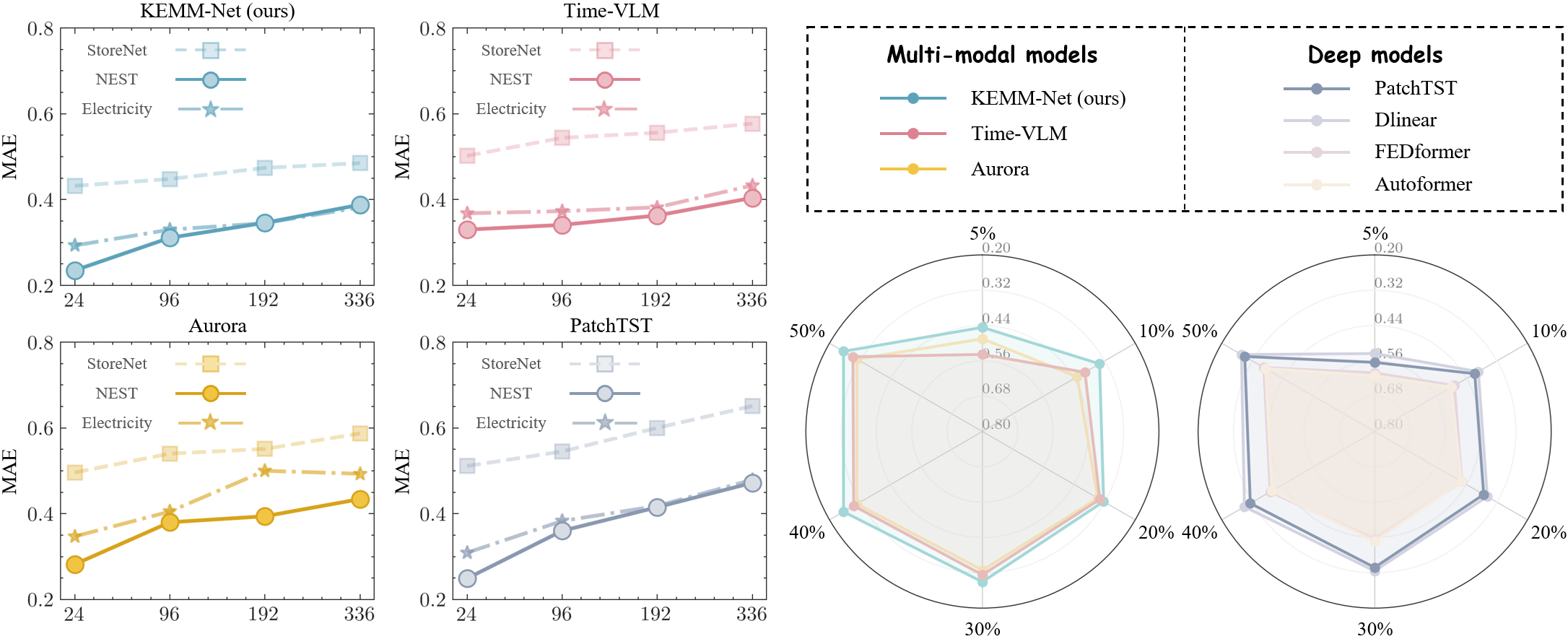}
\caption{Few-shot results. Left: Prediction MAE of different methods. Right: Radar plot: KEMM-Net excels under different data scarce level.}
\label{fig:fsl}
\vspace{-0.8em}
\end{figure*}

Following the standard setting~\cite{jintime,zhongtime,chenvisionts}, only 10\% training data are used. Fig. \ref{fig:fsl} presents the prediction results. 

\begin{itemize}[leftmargin=9pt]
  \item We select strong baselines for few-shot scenarios; among them, KEMM-Net demonstrates best performance, with only a slight increase in MAE. This is primarily due to its strong ability to integrate multi-modal knowledge. When available data is limited, the domain knowledge provided by text modality becomes increasingly essential. Through multi-modal knowledge alignment training, KEMM-Net is able to leverage this valuable information.
  \item The radar plots show different data availability from 5\% to 50\%, and MAE increases across all models. This phenomenon is particularly pronounced in traditional deep models. From the training logs, we observed that PatchTST's loss dropped to 0.22, yet the MAE on the test set exceeded 0.4, indicating that these deep models are prone to severe overfitting in few-shot scenarios. In contrast, KEMM-Net exhibits consistent robustness in few-shot setting, achieving the best results across various data availability levels. 
\end{itemize}

\begin{figure}[t]
\centering
\includegraphics[width=0.95\columnwidth]{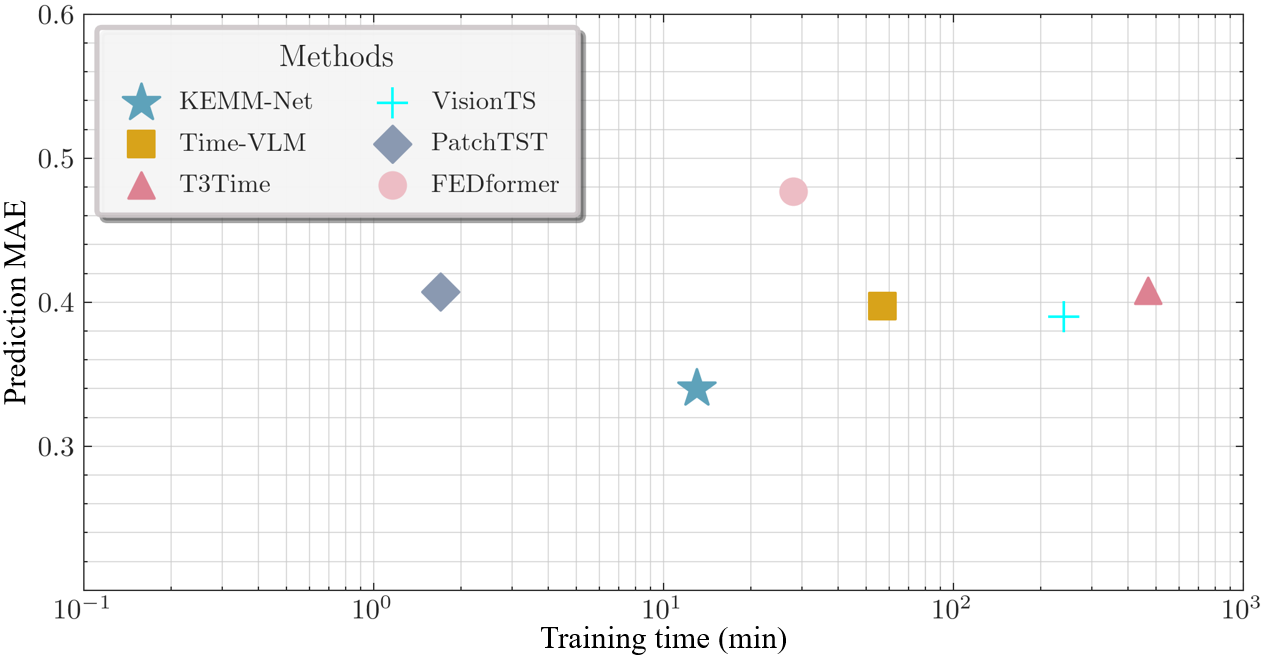}
\caption{MAE and training time of Electricity under few-shot setting}
\label{fig:compass}
\vspace{-0.8em}
\end{figure}

Figure \ref{fig:compass} shows the MAE and training time of different approaches under few-shot setting. Despite the substantial reduction in sample size, the two additional multi-modal models (we didn't include Aurora because it is a pre-trained foundation model), Time-VLM and T3Time, still consume considerable time for model training. On the other hand, while PatchTST has the shortest training time, its MAE increases significantly under few-shot setting. In contrast, KEMM-Net strikes an optimal balance between predictive performance and training efficiency.

\vspace{-0.3em}
\subsection{Discussion and Ablation}

\begin{figure*}[!t]
\centering
\includegraphics[width=1.0\textwidth]{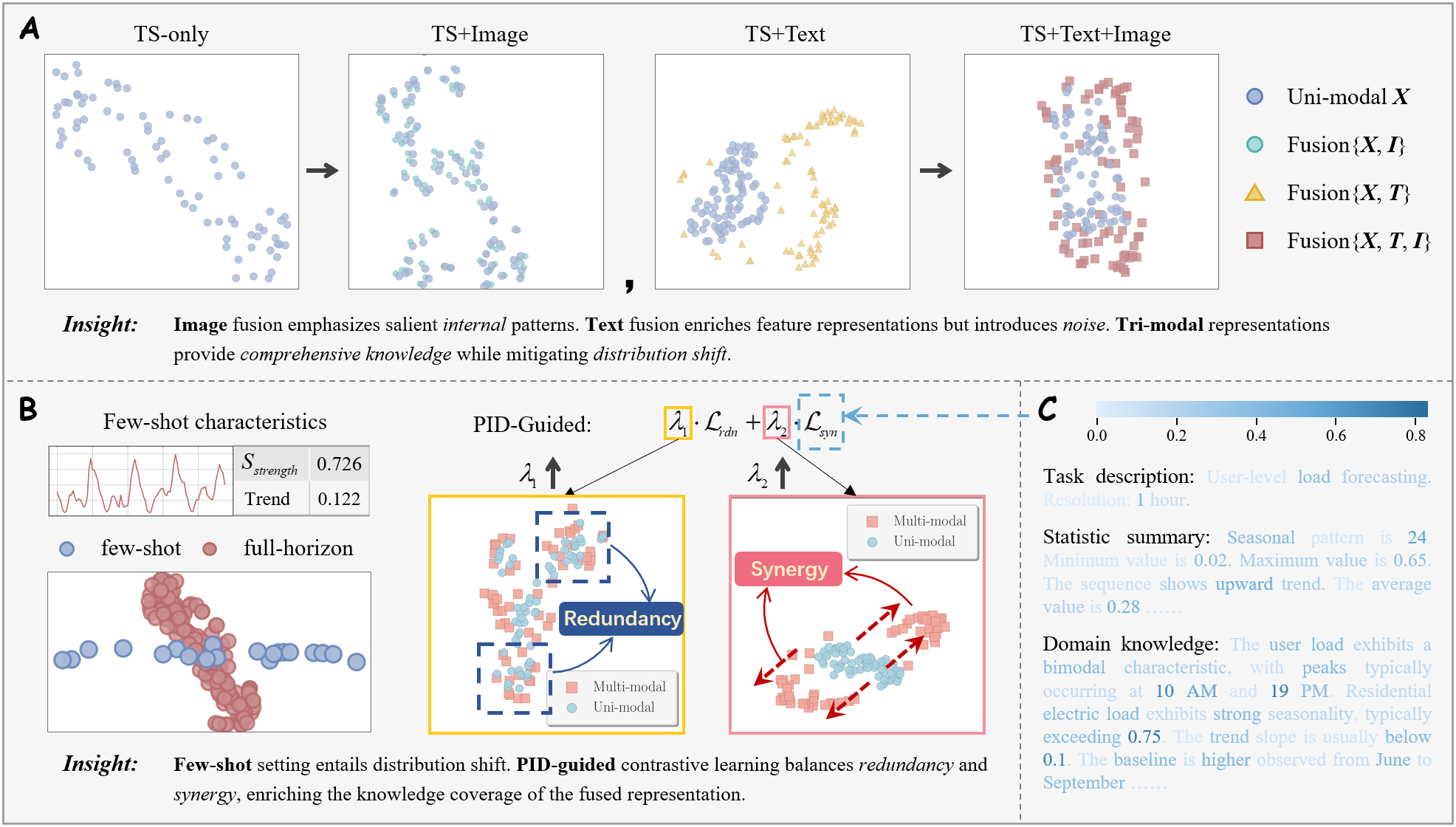}
\caption{Case study. A. t-SNE plot of unimodal \& fused representation; B. PID-guided alignment ; C. Token attention visualization}
\label{fig:case_study}
\vspace{-1em}
\end{figure*}

KEMM-Net leverages PID-guided multi-modal contrastive learning to align domain-specific semantics across modalities. We present a case study on the Electricity dataset (Fig. \ref{fig:case_study}).

\begin{itemize}[leftmargin=9pt]
  \item \textbf{Multi-modal augmentation yields richer knowledge representations.} As shown in Part A, we visualize via t-SNE the embeddings of: i) unimodal time series, ii) time series + image fusion, iii) time series + text fusion, and iv) tri-modal fusion. Image fusion emphasizes salient internal patterns, helping preserve prototypical characteristics such as daily periodicity. Text fusion enriches the representation, yet the overall distribution becomes overly dispersed, suggesting the introduction of noise. And finally, the tri-modal representation provides more comprehensive knowledge while avoiding distribution shift.
  \item \textbf{PID guidance balances redundancy and synergy.} Semantic alignment of domain knowledge is essential to fully exploit multi-modal information. Under few-shot settings, distribution shift limits the coverage of full-range temporal patterns, as shown in Part B. PID-guided contrastive learning explicitly controls redundancy and synergy: redundancy is reflected by the overlap between multi-modal and unimodal embeddings, whereas increasing synergy tends to ``push'' the multi-modal representation outward, thereby expanding knowledge coverage. It is worth noting that, in our view, redundancy should not be entirely suppressed, as it helps capture fine-grained features inherent in time series. The key to accurate forecasting lies in balancing these two characteristics, with details provided in Appendix C.
  \item \textbf{Unlocking synergy from domain knowledge.} From Parts A and B, the direction in which text expands unimodal embeddings mirrors the effect of the ${{\lambda }_{2}}$ term in PID guidance, suggesting that synergy is largely contributed by the text modality. To validate this hypothesis, we visualize token-level attention weights from the encoder following GPT-2, as shown in Part C. The model attends strongly to numerical tokens; moreover, it effectively leverages content in the \textit{domain knowledge} (e.g., “seasonality exceeding 0.75”), which differs from the few-shot estimate of 0.726 (Part B). Such externally supplied domain cues synergize with time series features to support downstream forecasting.
\end{itemize}

\begin{table}[t]
\centering
\caption{Ablation study on model design}
\label{tab:ablation}
\setlength{\tabcolsep}{7pt}
\resizebox{\columnwidth}{!}{
\begin{tabular}{l|cc|cc|cc}
\toprule
& \multicolumn{2}{c|}{StoreNet} & \multicolumn{2}{c|}{NEST} & \multicolumn{2}{c}{Electricity} \\
\cline{2-7}

& Full & FSL & Full & FSL & Full & FSL \\
\midrule

KEMM-Net & \textbf{0.384} & \textbf{0.461} & \textbf{0.243} & \textbf{0.323} & \textbf{0.251} & \textbf{0.340} \\
\midrule
w/o Text & 0.390 & 0.545 & 0.252 & 0.366 & 0.263 & 0.417 \\
$\rightarrow$w/o Stat. & 0.390 & 0.485 & 0.249 & 0.336 & 0.239 & 0.361 \\
$\rightarrow$w/o Know. & 0.386 & 0.527 & 0.246 & 0.358 & 0.233 & 0.379 \\
use word2vec & 0.389 & 0.525 & 0.250 & 0.359 & 0.258 & 0.405 \\
\midrule
w/o Image & 0.388 & 0.464 & 0.250 & 0.335 & 0.262 & 0.356 \\
GRU$\rightarrow$Transformer & 0.384 & 0.461 & 0.244 & 0.324 & 0.251 & 0.340 \\
\midrule
w/o CL & 0.402 & 0.530 & 0.264 & 0.359 & 0.273 & 0.404 \\
\bottomrule
\end{tabular}
}
\vspace{-1.0em}
\end{table}

We conduct ablation studies on the main architectural components, as summarized in Table \ref{tab:ablation}. Removing any single modality degrades forecasting performance. Within the textual modality, statistics have a larger impact in the full-shot setting, whereas domain knowledge is more critical in the few-shot setting. The effect of image is relatively consistent across both regimes. We also replace the GRU in the image embedding module with a Transformer and observe no clear improvement, indicating that a lightweight GRU is sufficient to model local pattern across only a few dozen frames. When removing PID-guided contrastive learning, the model struggled in capturing multi-modal semantics, and the MAE in the few-shot setting increases significantly, despite the presence of auxiliary information in other modalities.

\begin{table}[t]
\centering
\caption{Using same text on other multi-modal models}
\label{tab:text}
\setlength{\tabcolsep}{7pt}
\resizebox{\columnwidth}{!}{
\begin{tabular}{l|cc|cc|cc}
\toprule
& \multicolumn{2}{c|}{StoreNet} & \multicolumn{2}{c|}{NEST} & \multicolumn{2}{c}{Electricity} \\
\cline{2-7}

& Full & FSL & Full & FSL & Full & FSL \\
\midrule

Time-LLM & 0.444 & 0.567 & 0.288 & 0.372 & 0.287 & 0.423 \\
T3Time & 0.410 & 0.604 & 0.275 & 0.397 & 0.301 & 0.446 \\
Time-VLM & 0.429 & 0.545 & 0.262 & 0.360 & 0.264 & 0.389 \\

\bottomrule
\end{tabular}
}
\vspace{-1.0em}
\end{table}

Table \ref{tab:text} reports the results of applying the same text inputs used in KEMM-Net to other multi-modal models. Compared with Table I, these models show marginal gains in the full-shot setting, and even higher MAE on StoreNet. In the few-shot setting, Time-LLM and Time-VLM improve slightly, but still far less than KEMM-Net. These results indicate that simply adding extra information is insufficient and may introduce noise, further verifying the effectiveness of our domain semantic alignment mechanism.

\vspace{-0.3em}
\section{Conclusion}

In this paper, we study load time series forecasting in the energy industry and identify a key insight: the few-shot capability of multi-modal time series models is limited by inherent information redundancy rather than embedding or fusion learning. To overcome this, we propose KEMM-Net, which introduces domain prior knowledge during modal augmentation and leverages information-theoretic knowledge alignment to fully unlock multi-modal potential. Comprehensive experiments demonstrate that KEMM-Net achieves state-of-the-art performance.

\bibliographystyle{IEEEtran}
\bibliography{References}


 




\vspace{-0.5em}
\section*{Appendix}

\subsection{Proof for Corollary 1}

\renewcommand{\thefigure}{A\arabic{figure}}
\setcounter{figure}{0}
\renewcommand{\theequation}{A\arabic{equation}}
\setcounter{equation}{0}
\renewcommand{\thetable}{A\arabic{table}}
\setcounter{table}{0}

We establish the following assumption:

\begin{figure}[h]
\centering
\includegraphics[width=0.80\columnwidth]{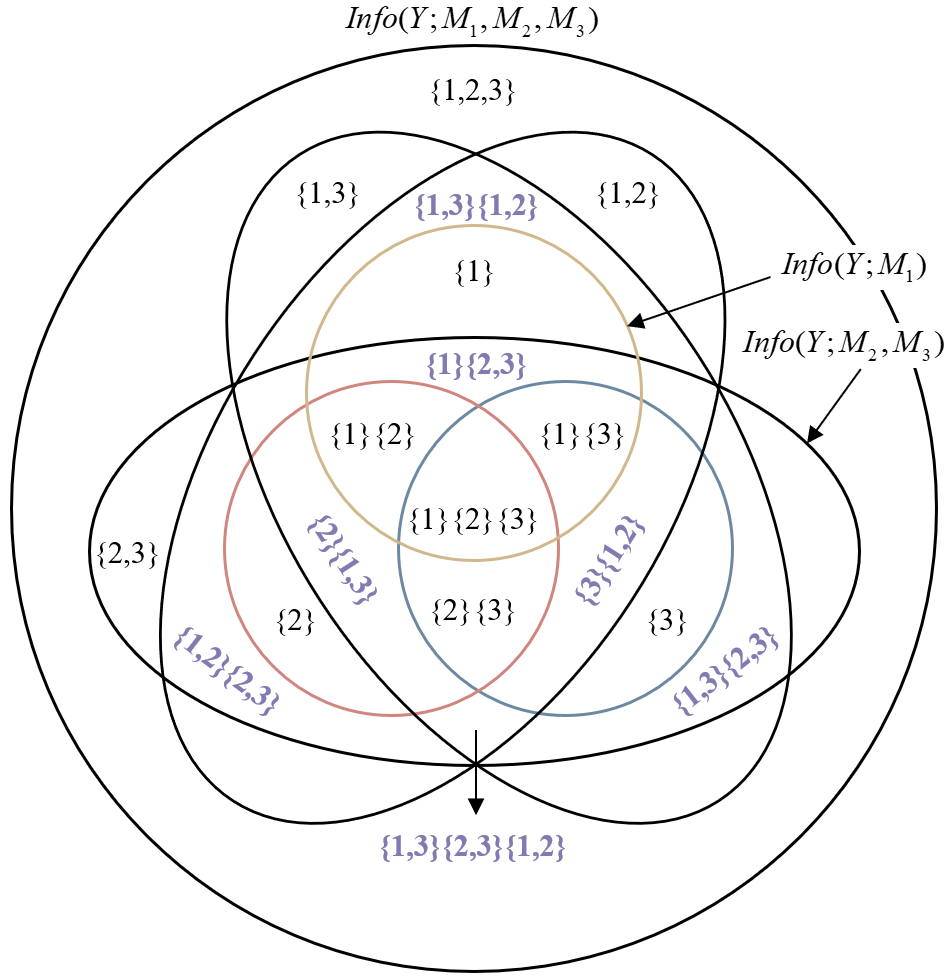}
\caption{PID for a four-variable system. $\{i,j\}$ represents the synergy between information $i$ and $j$, while $\{i\}\{j\}$ represents their redundancy. All entangled terms are highlighted in purple bold.}
\label{fig:A1}
\vspace{-0.5em}
\end{figure}

\noindent
\textbf{Assumption 3.} When modeling the total information contribution from three modalities, the interactions involving the modality $X$ are retained, while the interactions between $T$ and $I$ are neglected. As a result, the original four-variable system (Fig. A1) degenerates to the form shown in Fig. A2.

Although this assumption may seem strong, it is reasonable in our knowledge enrichment context. Both text $I$ and image $T$ modalities are used to enhance the representation of the time series, where the $I$ modality is derived from a transformation of the time series while $T$ introduces domain prior knowledge beyond the raw numerical observations. Therefore, from the perspective of knowledge synergy, the semantic connection between $T$ and $I$ is inherently weaker (since text branch mainly attends to \textit{domain-knowledge} tokens. See Fig. 7. C). 

\begin{figure}[t]
\centering
\includegraphics[width=0.80\columnwidth]{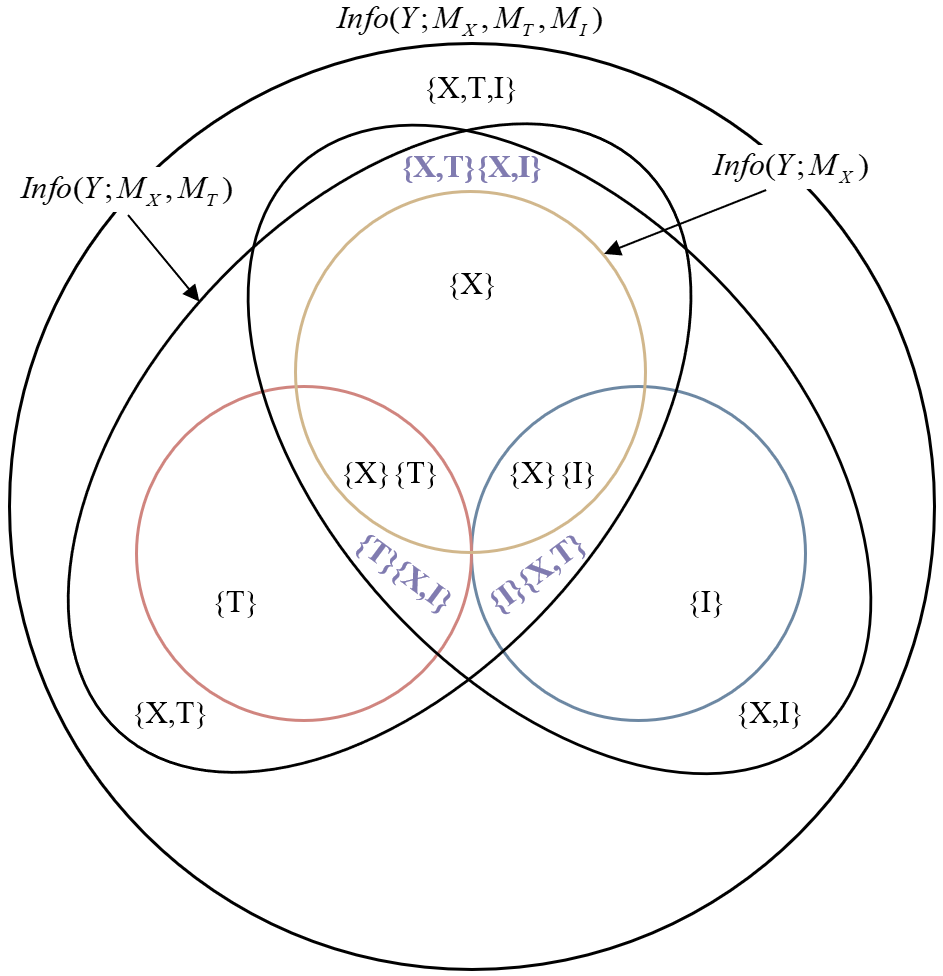}
\caption{Simplified system for three modalities. In this form, many of the interactions between the text and image modalities are omitted, while all terms involving modality $X$ are retained.}
\label{fig:A2}
\vspace{-1.0em}
\end{figure}

In this term, when we calculate the information contribution from $X$ and $T$, denoted as $Info(Y;{{M}_{X}},{{M}_{T}})$ , the terms we consider include $\{X\}$, $\{X\}\{I\}$, $\{X\}\{T\}$, $\{T\}$, $\{X,T\}$, and the entangled terms. The terms $\{X\}$, $\{X\}\{I\}$, and $\{X\}\{T\}$ can be combined into $Info(Y;{{M}_{X}})$, as shown by the yellow circle in the figure. $\{T\}$ corresponds to $U_T$, and $\{X,T\}$ corresponds to $S_{XT}$. Therefore, we obtain the following:

\vspace{-1.0em}
\begin{equation}\label{eq:x}
Info(Y;{{M}_{X}},{{M}_{T}})={{S}_{XT}}+{{U}_{T}}+Info(Y;{{M}_{X}})+E{{I}_{XT}}
\end{equation}

Corollary 1 is proven.

\vspace{-0.3em}
\subsection{Datasets Details}

\renewcommand{\thefigure}{B\arabic{figure}}
\setcounter{figure}{0}
\renewcommand{\theequation}{B\arabic{equation}}
\setcounter{equation}{0}
\renewcommand{\thetable}{B\arabic{table}}
\setcounter{table}{0}

\begin{table}[h] 
\vspace{-0.8em}
    \centering 
    \caption{Details of the datasets}
    \label{tab:dataset} 

    \setlength{\tabcolsep}{6pt} 
    \resizebox{\columnwidth}{!}{ 
    \begin{tabular}{ c >{\centering\arraybackslash}m{5.0cm} c c c}
        \toprule
        \textbf{Datasets} & \textbf{Description} & \textbf{Dim.} & \textbf{Resolution} & \textbf{Steps} \\ 
        \midrule
        Electricity & User-level load data. Standard benchmark & 321 & 1 hour & 17520 \\
        \midrule
        Nest & Building multi-energy load dataset & 8 & 15 min & 35040 \\
        \midrule
        StoreNet & Distribution network load dataset & 5 & 15 min & 35040 \\
        \bottomrule
    \end{tabular}
    }
    \vspace{-1.0em}
\end{table}

\vspace{-0.5em}
\subsection{More Ablation and Hyperparameter Sensitivity Analysis}

\renewcommand{\thefigure}{C\arabic{figure}}
\setcounter{figure}{0}
\renewcommand{\theequation}{C\arabic{equation}}
\setcounter{equation}{0}
\renewcommand{\thetable}{C\arabic{table}}
\setcounter{table}{0}

\noindent
\textbf{Redundancy and synergy loss ratio}. The model achieves its best performance when the ratio ${{\lambda }_{1}}:{{\lambda }_{2}}$ is set to 1:3. As the weight of ${{\lambda }_{1}}$ increases, the model overemphasizes redundancy and degrades performance, particularly in few-shot scenarios. Conversely, an overly large ${{\lambda}_{2}}$ slightly increases prediction error, suggesting that excessive noise is introduced and weakens the learning of typical temporal patterns.

\noindent
\textbf{External-source information input.} We fed sudden weather news, policy reports into the instruction-finetuned LLM, which extracts load-related cues and incorporates them into the domain knowledge field. As shown in Table C1, such external inputs bring slight performance gains, validating the compatibility of KEMM-Net with external-source information. Due to space constraints, the complete external inputs will be provided in the second-round review.

\begin{figure}[t]
\centering
\includegraphics[width=1.0\columnwidth]{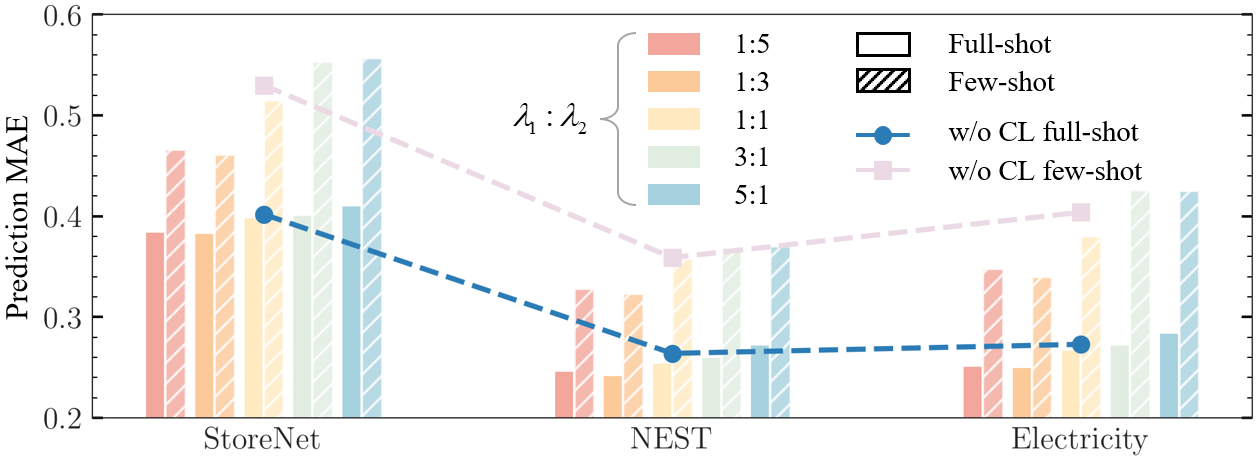}
\caption{Impact of the choice of ${{\lambda }_{1}}$ and ${{\lambda }_{2}}$}
\label{fig:D1}
\vspace{-0.8em}
\end{figure}

\begin{table}[h]
\centering
\caption{Using external source information input}
\label{tab:text}
\setlength{\tabcolsep}{7pt}
\resizebox{\columnwidth}{!}{
\begin{tabular}{l|cc|cc|cc}
\toprule
& \multicolumn{2}{c|}{StoreNet} & \multicolumn{2}{c|}{NEST} & \multicolumn{2}{c}{Electricity} \\
\cline{2-7}

& Full & FSL & Full & FSL & Full & FSL \\
\midrule

KEMM-Net & 0.384 & 0.461 & 0.243 & 0.323 & 0.251 & 0.340 \\
+ external info. & 0.372 & 0.458 & 0.233 & 0.319 & 0.240 & 0.335 \\

\bottomrule
\end{tabular}
}
\vspace{-0.5em}
\end{table}

\vspace{0.2em}
\noindent
\textbf{More parameter implementation details.} For image resolution, we adopt an input size of $224\times 224$; reducing the resolution to $128\times 128$ and $56\times 56$ increases the MAE by 0.8\% and 1.2\%, respectively. While image size has a relatively minor impact on overall performance, we retain $224\times 224$ to remain consistent with the CLIP pretraining setting. For negatives construction in image branch, we observed that the number of crops has little impact on performance. For color jittering, we set the random perturbation magnitude to 0-75\%, since overly small transformations slow down the decrease of the contrastive loss. The instruction-finetuned LLM used in this study is Llama3-70B, and we found no significant effect on the final results by using other LLMs such as gpt.

\subsection{Example of Instruction Finetune Prompt}

\renewcommand{\thefigure}{D\arabic{figure}}
\setcounter{figure}{0}
\renewcommand{\theequation}{D\arabic{equation}}
\setcounter{equation}{0}
\renewcommand{\thetable}{D\arabic{table}}
\setcounter{table}{0}

\vspace{0.3em}
\begin{promptbox}
\promptrole{System}

\vspace{0.8em}
\phead{Setup}
You are an intelligent agent specializing in time series analysis.
Given a user-provided load sequence, summarize it from three aspects:
task description, statistical summary, and domain knowledge.
For domain knowledge, follow the reasoning steps below.

\vspace{0.8em}
\phead{Chain-of-thought}
\hlone{Step 1}: Identify the load type.

\hltwo{Step 2}: Analyze potential behavior patterns and how they affect electricity demand.

\hlthree{Step 3}: Describe the long-term variation trend of this load type.

......

\vspace{0.8em}
\vspace{0.5mm}
\promptrole{User}

\{``load'': [0.326, 0.339, 0.347, 0.585, 0.660, \ldots]\}

\vspace{0.5mm}
\vspace{0.8em}
\promptrole{Assistant}

Task Description: User-level load forecasting task...

Statistic Summary: The maximum value is 0.65. The minimum value is 0.02. The seasonal pattern is 24. The average value is 0.28 ...

Domain Knowledge: \hlone{The user load exhibits a bimodal characteristic,
with peaks typically occurring at 10 AM and 19 PM.} \hltwo{Residential electric load exhibits strong seasonality, typically exceeding 0.75. The trend slope is usually below 0.1.} \hlthree{The baseline load is higher
from June to September}...

\end{promptbox}

\vfill

\end{document}